\documentclass{aa}  
\usepackage{multirow}
\usepackage[shortlabels]{enumitem}
\usepackage{ulem}
\usepackage{txfonts}
\usepackage[%
  breaklinks = true,
  colorlinks = true,
  urlcolor   = blue,
  citecolor  = blue,
  linkcolor  = blue,
]{hyperref}
\graphicspath{ {./figures/} }

\newcommand{\IRIS}{\textit{IRIS}}
\newcommand{\SDO}{\textit{SDO}}

\newcommand{\angstrom}{\textup{\AA}}
\newcommand{\Halpha}{{H$\alpha$}}
\newcommand{\Hbeta}{{H$\beta$}}
\newcommand{\CaII}{{\ion{Ca}{II}}}
\newcommand{\SiIV}{{\ion{Si}{IV}}}
\newcommand{\CII}{{\ion{C}{II}}}

\newcommand{\OI}{{\ion{O}{I}}}
\newcommand{\OIV}{{\ion{O}{IV}}}

\newcommand{\MgII}{{\ion{Mg}{II}}}

\newcommand{\Heteneightthirty}{{\ion{He}{I}  $10830$\,$\angstrom$}}

\newcommand{\vlos}{{$V_{\mathrm{LOS}}$}}

\newcommand{\kmeans}{{$k$-means}}

\begin{document} 

   \title{Solar surges related to UV bursts}

   \subtitle{Characterization through \kmeans, inversions and density diagnostics}

   \author{D. N\'obrega-Siverio\inst{1,2,3,4}
          \and
          S.L. Guglielmino\inst{5,6}
          \and
          A. Sainz Dalda\inst{7,8,9}
          }

   \institute{Instituto de Astrof\'isica de Canarias, E-38205 La Laguna, Tenerife, Spain\\
   \email{dnobrega@iac.es}
             \and
             Universidad de La Laguna, Dept. Astrof\'isica, E-38206 La Laguna, Tenerife, Spain
             \and
             Rosseland Centre for Solar Physics, University of Oslo, PO Box 1029 Blindern, 0315 Oslo, Norway
             \and
             Institute of Theoretical Astrophysics, University of Oslo, PO Box 1029 Blindern, 0315 Oslo, Norway\\
             \and
              Dipartimento di Fisica e Astronomia ``Ettore Majorana'' -- Sezione Astrofisica, \\Universit\`a degli Studi di Catania, Via S. Sofia 78, I-95123 Catania, Italy
              \and
              INAF -- Osservatorio Astrofisico di Catania, Via S.~Sofia 78, I-95123 Catania, Italy \\
              \and
              Lockheed Martin Solar \& Astrophysics Laboratory, Palo Alto, CA 94304, USA
              \and 
              Bay Area Environmental Research Institute, NASA Research Park, Moffett Field, CA 94035, USA
              \and
              Stanford University, HEPL, 466 Via Ortega, Stanford, CA 94305-4085, USA
             }

   \date{Received June 7, 2021; accepted August 31, 2021}

%
%
 
  \abstract
    {Surges are cool and dense ejections typically observed in chromospheric lines and closely related to other solar phenomena like UV bursts or coronal jets. Even though surges have been observed for decades now, questions regarding their fundamental physical properties such as temperature and density, as well as their impact on upper layers of the solar atmosphere remain open.}
   {Our purpose is to address the current lack of inverted models and diagnostics of surges, as well as characterizing the chromospheric and transition region plasma of these phenomena.}
   {We have analyzed an episode of recurrent surges that appear related to UV bursts observed with the \textit{Interface Region Imaging Spectrograph} (\IRIS) on 2016 April. 
   The mid- and low-chromosphere of the surges are unprecedentedly examined by getting their representative \MgII\ h\&k line profiles through the \kmeans\ algorithm and performing inversions on them using the state-of-the-art STiC code. 
   We have studied the far-UV spectra focusing on the \OIV\ 1399.8~\AA{} and 1401.2~\AA{} lines, unexplored for surges, carrying out density diagnostics to determine the transition region properties of these ejections. We have also used numerical experiments performed with the Bifrost code for comparisons.}
   {
   Thanks to the \kmeans\ clustering, we reduce the number of \MgII\ h\&k profiles to invert by a factor 43.2.
   The inversions of the representative profiles show that the mid- and low-chromosphere of the surges are characterized, with a high degree of reliability, by temperatures mainly around T $=6$~kK at $ -6.0 \le \log_{10}(\tau) \le -3.2$. For the electronic number density, $n_e$, and line-of-sight velocity, \vlos, the most reliable results from the inversions are within $ -6.0 \le \log_{10}(\tau) \le -4.8$, with $n_e$ ranging from $\sim1.6 \times 10^{11}$ cm$^{-3}$ up to $10^{12}$ cm$^{-3}$, and \vlos\ of a few km s$^{-1}$.
   We find, for the first time, observational evidence of enhanced \OIV\ emission within the surges, indicating that these phenomena have a considerable impact on the transition region even in the weakest far-UV lines. The \OIV\ emitting layers of the surges have an electron number density ranging from $2.5\times 10^{10}$ cm$^{-3}$ to $10^{12}$ cm$^{-3}$. 
   The numerical simulations provide theoretical support in terms of the topology and of the location of the \OIV\ emission within the surges.}
   {}


   \maketitle

%
%
\section{Introduction}\label{sec:introduction}

Within the catalog of solar eruptive phenomena, surges are key chromospheric plasma ejections due to their frequent appearance as well 
as their close relationship with other important scientific targets of the Sun. Most common examples are 
%
flux emergence regions \citep[e.g.,][]{Brooks2007,Guglielmino:2010lr,Yang:2018,Guglielmino:2018,
Guglielmino:2019,Kontogiannis:2020,Verma:2020},
Ellerman bombs \citep[e.g.,][]{watanabe2011,Yang:2013h,Vissers:2013,Rutten:2013}, 
UV bursts \citep[e.g.,][]{Kim:2015y,Huang:2017z,Rouppe2017,Ortiz:2020},
coronal jets \citep[e.g.,][]{Yokoyama:1996kx,canfield1996,Zhang2014,Nelson:2019,Ruan:2019,Joshi:2020}, 
and flares \citep[e.g.,][]{McMath:1948,Tandberg-Hanssen:1959,roy1973,Wang:2012,Huang:2014,Schrijver:2015},
although there are also evidences of surges or surge-like ejections associated with light bridges \citep{Asai2001,Shimizu:2009,Robustini:2016,Tian:2018}, explosive events \citep{Madjarska2009}, and eruptive filaments \citep{Sterling:2016,Li:2017}. 

Mainly observed in \Halpha, and in other lines such as \CaII \ 8542 $\angstrom$, \Heteneightthirty, \Hbeta\ 4861 $\angstrom$ and 
\CaII\ H \& K \citep{Zhang:2000hb,Liu2004,Nishizuka:2008zl,liu2009,Vargas2014}, it is now established, from both observations and numerical experiments, that 
surges are ejections typically with characteristic lengths of $10 - 50$ Mm 
(sometimes even reaching $200$~Mm), consist of 
small-scale thread-like features \citep{nelson2013, Li2016}, can be seen as a dense wall-like structure that appears surrounding 
the emerged region \citep{Moreno-Insertis:2008ms, Moreno-Insertis:2013aa}, and their ejection direction depends on the particular 
geometry of the ambient magnetic field \citep{MacTaggart:2015}. In addition, they can be recurrent \citep{Jiang2007,Uddin2012,Wang2014} 
and/or have oscillations, as well as rotational and helical motions \citep{Jibben2004,Jiang2007,Bong:2014,Cho:2019}.

Despite this observational and theoretical effort, important questions concerning surges remain open. For instance, observational studies have mainly focused on their kinematic properties \citep[see, e.g.,][]{roy1973,Gu:1994, Chae1999,Chen2008,Ortiz:2020}, so fundamental quantities such as density and temperature, and their distribution with height are not known. Recent simulations by \cite{Nobrega-Siverio:2016} using the Bifrost code \citep{Gudiksen:2011qy} point out that surges are composed by multi-thermal plasma populations that result from efficient heating/cooling mechanisms during their ejection. To provide observational support to these findings, it is necessary to properly derive constraints from chromospheric lines whose formation occurs in non-local thermodynamic equilibrium (NLTE) conditions and, moreover, depend on radiative transfer effects such as partial frequency redistribution and 3D scattering (see \citealp{Leenaarts:2012prd}): a complex task that is still pending. Another key aspect that remains poorly known is the impact of the surges on the transition region and corona. There are only a few examples in the literature where the relationship between transition region lines and surges is studied, suggesting that a thin hot layer may surround the cooler \Halpha\ surge \citep[see, e.g.,][]{Kirshner1971,Schmieder1984,Madjarska2009}. 
Combining theory and observations, \cite{Nobrega-Siverio:2017,Nobrega-Siverio:2018} show that surges have enhanced emission on \SiIV\ that is related to nonequilibrium ionization effects during the surge formation; nonetheless, the \IRIS\ \citep{De-Pontieu:2014iris} observations analyzed by these authors only have spectral information at the base of the surge due to the location of the raster. 
Independent observational evidence of enhanced \SiIV\ emission associated with surges with \IRIS\ is found 
by \cite{Guglielmino:2019}. Nevertheless, none of the aforementioned authors have performed diagnostics to extract valuable information of the hottest regions of the surges and the weakest lines of the transition region are still unexplored.

The purpose of this work is to 
tackle the present lack of inverted models and diagnostics of surges, filling the gap in the knowledge of important physical quantities of these key phenomena for the solar atmosphere. To that end, we analyze a series of surges observed by \IRIS\ within Active Region (AR) NOAA 12529 \citep{Guglielmino:2017,Guglielmino:2019umbra}. For the first time, characterization of the chromospheric and transition region properties of the surge plasma is achieved combining machine learning techniques, inversions, and density diagnostics. In addition, realistic simulations are used to provide theoretical support to the observations. 
The layout of this paper is as follows. In Section \ref{sec:obs_and_methods}, we describe the observational data, the methods, and the numerical experiments employed. In Section \ref{sec:results}, we study the surges focusing on the \MgII\ h\&k line (Section \ref{sec:mgii_analysis}), the \OIV\ lines (Section \ref{sec:oiv_analysis}) and the comparisons with the simulations (Section \ref{sec:experiment_analysis}). Finally, Section \ref{sec:discussion} contains the discussion and main conclusions.

%
%
\section{Observations, methods, and numerical experiments}\label{sec:obs_and_methods}

\subsection{Observational data}\label{sec:observations}
 For this paper, we have used \IRIS\ observations obtained on 2016 between 22:34:43 UT on April 13 and 01:55:29 UT on April 14, within the active region AR NOAA 12529, which passed on the solar disk exhibiting an almost typical bipolar $\beta$ configuration, at heliocentric coordinates (X,Y) $\approx (0\arcsec, 260\arcsec)$ (central solar meridian passage) and heliocentric angle $\mu=0.96$. In particular, we have focused on 
 the recurrent surges that appear as a consequence of magnetic flux emergence. The flux emergence episode, originated in the diffuse plage of the following positive polarity of the AR, as well as the long-lasting (3 hours) UV bursts have been analyzed in detail by \cite{Guglielmino:2018,Guglielmino:2019}.
 
 The \IRIS\ data set corresponds to the program OBS3610113456 whose sequence consists of six dense 64-step raster scans. In this paper we have employed a more recent calibration (version 1.84) of the \IRIS\ level 2 data products than the one originally applied by \citet{Guglielmino:2018,Guglielmino:2019} (version 1.56) due to updates carried out by the instrument team. In addition,  we have only used the last four scans of the sequence due to a change on the exposure time during the observations, maintaining the notation previously used to identify them, namely, Raster 3, 4, 5 and 6. The selected raster scans contain UV spectra acquired in seven spectral ranges, encompassing \CII~1334.5 and 1335.7~\AA{}, \SiIV~1394 and 1402~\AA{}, \MgII~k~2796.3 and h~2803.5~\AA{} lines and the faint lines around chromospheric \OI~1355.6~\AA{} line. The exposure time is 18~s for the far-UV (FUV) channels and to 9~s for the near-UV (NUV) channel. The sequence has a pixel size of 0\farcs35 along the \textit{y} direction, with a 0\farcs33 step size along the \textit{x} direction, and the covered field of view (FoV) was $22\farcs3 \times 128\farcs4$. Slit-jaw images (SJIs) were simultaneously acquired in the 1400 and 2796~\AA{} passbands, with a cadence of 63~s. In addition, to link the \IRIS\ observations to the photospheric magnetic field, we have employed data from the \textit{Helioseismic and Magnetic Imager} \citep[HMI;][]{Scherrer:2012qf} on board the \textit{Solar Dynamics Observatory} \citep[SDO;][]{Pesnell:2012}. For further details about the \IRIS/\SDO\ data set, we refer the reader to \citet{Guglielmino:2018,Guglielmino:2019}.

\begin{figure*}[!t]
	\centering
    \includegraphics[width=1.0\textwidth]{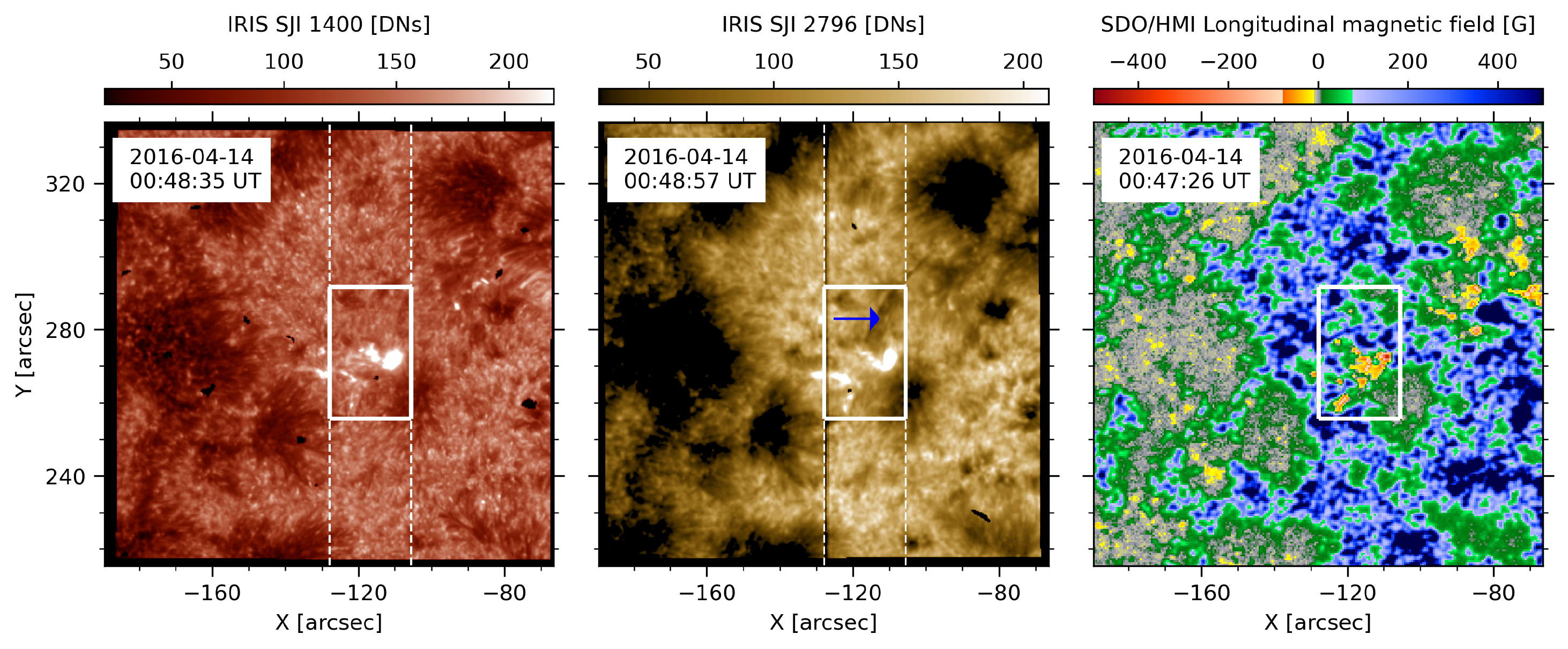}
	\caption{Context image of the events associated with the flux emergence region  in the plage following the positive polarity of the AR NOAA 12529 at $\sim$00:48 UT on 2016 April 14. 
	Left: \IRIS\ SJI 1400~\AA{}. Middle: \IRIS\ SJI 2796~\AA{}. Right: \SDO/HMI line-of-sight (LOS) magnetogram. The vertical dashed lines indicate the area covered by the \IRIS\ slit during  the raster scans. The solid box, ranging from X=-127\farcs9 to X=-105\farcs6 and Y=255\farcs7 to Y=291\farcs6, frames the portion of the FoV studied in this paper. The blue arrow indicates one of the observed surges. The axes give the distance from solar disk center. \label{fig_context}}
\end{figure*}

Figure~\ref{fig_context} displays the context of the observations at $\sim$00:48 UT on 2016 April 14. The left panel contains the \IRIS\ SJI 1400~\AA{}, showing a strong brightening (a UV burst) located around $X=-110\arcsec$ and $Y=272\arcsec$. In the middle panel, which illustrates the \IRIS\ SJI 2796~\AA{}, the UV burst is visible next to an elongated dark structure that corresponds to a surge (blue arrow). The right panel contains the SDO/HMI line-of-sight (LOS) magnetogram, showing the diffuse plage that follows the positive polarity of the AR. At the time of the image, a flux emergence episode is taking place and a negative polarity patch is clearly distinguishable embed in the positive polarity dominated region. In the figure, the portion of the FoV analyzed in this paper is delimited by a solid box encompassing a subFoV of $22\farcs3 \times 35\farcs9$, which ranges from X=-127\farcs9 to X=-105\farcs6 and Y=255\farcs7 to Y=291\farcs6 at the time of the image.

\subsection{Methods}\label{sec:methods}

\subsubsection{\texorpdfstring{\kmeans}{k-means} clustering}\label{sec:kmeans}
\kmeans\ is an unsupervised clustering algorithm, popular in data-mining, machine learning, and artificial intelligence, that classifies a set of $n$ samples in $k$ disjoint groups (or clusters) of equal variance. To that end, it minimizes the within-cluster sum-of-squares (also known as inertia) given by
\begin{equation}
\sum_{i=1}^{n} \sum_{j=1}^{k} \min \left({ \left|\left| x_i - \mu_j \right|\right|^2}\right),
\end{equation}
where $x_i$ is the $i$-th observed point and $\mu_j$ the mean of the $j$-th cluster center or centroid. 

This algorithm is relatively easy to implement, scales to large data sets and can be adapted to group line profiles. In solar physics, \kmeans\ has gained popularity being used, for instance,  to classify Stokes profiles and chromospheric lines in different phenomena such as flares and spicules \citep[see, e.g.,][]{Pietarila:2007,Viticchie:2011,Panos:2018,Sainz-Dalda:2019,Bose:2019,Robustini:2019,JoshiJayant:2020,Kuckein:2020,Bose:2021,Bose:2021b,Barczynski:2021}. 

In our case, we have used it to classify the \MgII\ h\&k profiles obtained by \IRIS\ within the subFoV shown as a solid box in Figure \ref{fig_context}.
This way, the \kmeans\ method can help us to identify whether surges have characteristic \MgII\ h\&k line profiles that 
may indicate similar atmospheric properties. The reason to classify the spectra is inspired by the theoretical work by \cite{Nobrega-Siverio:2016}, 
in which the authors discern different populations within the surge depending on the thermodynamics and evolution of the plasma.
 Details about the data processing to feed the \kmeans\ algorithm as well as the choice of the number of clusters $k$ can be found in Appendix \ref{sec:appendix}.

\subsubsection{Inversions}\label{sec:inversions}

Inversions of the \MgII\ h\&k lines have been carried out using the STockholm inversion Code\footnote{\url{https://github.com/jaimedelacruz/stic}} \citep[STiC,][]{de-la-Cruz-Rodriguez:2019,de-la-Cruz-Rodriguez:2016}, which assumes NLTE and includes partial frequency redistribution effects of scattered electrons \citep{Leenaarts:2012prd}.  This code allows us to get a stratified model of the solar atmosphere that covers, depending on the spectral lines considered to be inverted, the photosphere, the chromosphere, and the transition region. An initial guess model atmosphere is slightly modified in an iterative process until the output radiation, obtained by solving the radiative transfer equation taking the considerations mentioned above into account, fits the observed spectra. The interested reader in the problem of the inversion of spectral data in the context of the solar physics can find an excellent review by \cite{del-Toro-Iniesta:2016}.

We also explored the possibility of inverting the \MgII\ h\&k surge profiles using \IRIS$^2$ \citep{Sainz-Dalda:2019}\footnote{\url{https://iris.lmsal.com/iris2/}}. Its inversion scheme (i.e., the number of nodes used in the thermodynamics variables to vary the model atmosphere at different optical depths) consists of two cycles: in the first cycle, it uses 4 nodes in temperature, and 3 nodes both in the line-of-sight and turbulent velocities; in the second cycle, it uses 7 nodes in temperature and 4 nodes in both velocities. Even though this approach is good enough for a large variety of profiles, it struggles with profiles with extreme line widths, such as those observed in X-class flares and UV bursts, and with profiles related to eruptive events. This means that the current version of \IRIS$^2$ is not appropriated for phenomena like surges.

\begin{table}
\centering
\begin{tabular}{ c c c c c}
 \hline
  \hline
 \rule{0pt}{2.5ex}
Nodes in variable for cycle \# & 1 & 2 & 3 & 4 \\
 \hline
\rule{0pt}{2ex} 
 $T$ & 4 & 7 & 9 & 9 \\ 
$V_{turb}$ & 3 & 4 & 7 & 9 \\  
\vlos\ & 3 & 4 & 4 & 6\\
\hline 
\end{tabular}
\caption{Number of nodes per cycle considered for the inversion of the \MgII\ h\&k spectral profiles in this article.\label{table:inversions}}
\centering
\end{table}

Taking the above into consideration, we have proceeded to invert the \MgII\ h\&k profiles of our observations with a more complex scheme than the one used to build the \IRIS$^2$ database. After different tests, we have selected an inversion scheme that consists of 4 cycles whose details are summarized in Table \ref{table:inversions}. For cycles 1 and 2, the number of nodes are the same as those used to build the $IRIS^2$ database. The temperature, $T$, and electron density, $n_e$, in the initial guess model atmosphere for the cycle 1 are taken from the FALC model \citep{Fontenla:1993}. The micro-turbulence velocity, $V_{turb}$, and the line-of-sight velocity, \vlos, are introduced {\it ad-hoc} as a smooth gradient along the optical depth between $-6 \le \log_{10}(\tau) \le 0$ that goes from 3 to 1 km s$^{-1}$ for $V_{turb}$, and from 1 to 0.5 km s $^{-1}$ for \vlos. The initial model atmosphere for the cycles 2, 3, and 4 is the output model atmosphere obtained from the previous cycle. The computation of the inversion uncertainties, $\sigma$, is carried out following Equation (42) of the paper by \cite{del-Toro-Iniesta:2016}: the same calculation performed by \cite{Sainz-Dalda:2019} for the \IRIS$^2$ database.

\subsubsection{Density diagnostics}\label{sec:density_diagnostics}

The density diagnostics method is based on the theoretical relationship between electron density and intensity ratio of the \OIV\ 1399.8~\AA{} and 1401.2~\AA{} lines. Indeed, these intercombination lines are known to provide useful electron density diagnostics in a variety of solar features and astrophysical plasmas \citep[see, e.g.,][]{Feldman:1979}, because their ratios are largely independent of the electron temperature, and only weakly dependent on the electron distribution \citep{Dudik:2014}. A recent discussion about the diagnostic potential of these lines in \IRIS\ observations has been presented by \citet{Polito2016}, who also showed that their sensitivity interval for density diagnostics range between $10^{10} - 10^{12} \,\mathrm{cm}^{-3}$.

The theoretical intensity ratios between the \OIV\ 1399.8~\AA{} and 1401.2~\AA{} lines were used to derive the electron density from the observed line ratio, under the assumption of ionization equilibrium  (CHIANTI v.8 database, \citealp{DelZanna:2015}), with formation temperature of $\log_{10}(T) = 5.15$~K. We obtained the line peak intensity from each considered pixel by fitting a Gaussian function to the \OIV\ 1399.8~\AA{} and 1401.2~\AA{} line profiles, respectively, also retrieving the uncertainties of the intensities. Blends in these emission lines are negligible in our study, as they mostly occur during flares \citep{Polito2016_flare}. Finally, for the computation of the density, we used the IDL \textit{SolarSoft} routine \texttt{iris}\textunderscore\texttt{ne}\textunderscore\texttt{oiv.pro}.

\begin{figure*}[ht]
\centering
\includegraphics[width=0.82\textwidth]{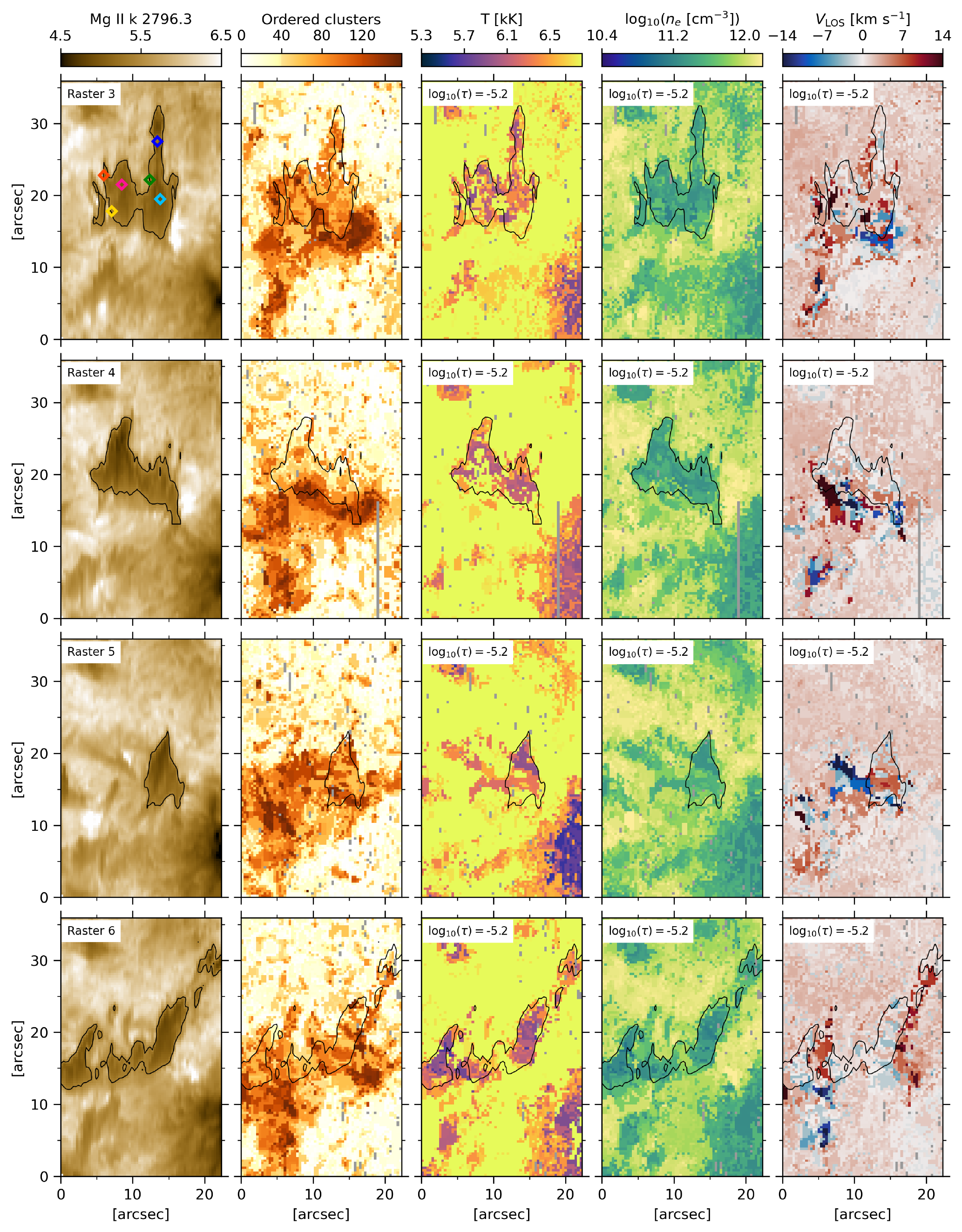}
\caption{Observed surges and results from the \kmeans\ and inversions. From left to right columns: radiance maps in the core of the \MgII\ k 2796.3~\AA{} line;  
cluster labels from the \kmeans\ ordered by the amount of profiles within a cluster; and maps at $\log_{10}(\tau) = -5.2$ for
temperature, $T$,  electron number density, $n_e$, and line-of-sight velocity, \vlos, 
from the inversions of the \MgII\ h\&k line using the STiC code.  
Black contours delimit the bulk of the surges visible in the \MgII\ k 2796.3~\AA{} radiance maps.
Bad pixels and pixels affected by cosmic rays are masked and shown with color grey.
Colored diamonds are superimposed in the first panel for later reference.
An animation of this figure is available varying $\log_{10}(\tau)$ from  $-6.0$ to $-3.2$.}
\label{fig_inversions_map}
\end{figure*}

\subsection{Numerical experiments}\label{sec:numerical_experiments}

To get a joint perspective between observations and theory, we have used three different 2.5D numerical experiments 
in which surges are ejected as a result of magnetic flux emergence. The simulations were performed with 
the radiation-magnetohydrodynamic Bifrost code \citep{Gudiksen:2011qy} assuming LTE and including thermal conduction along the magnetic field lines, optically thin losses, and radiation transfer adequate to the photosphere and chromosphere. The details of these simulations have been described in the papers by \cite{Nobrega-Siverio:2016,Nobrega-Siverio:2017,Nobrega-Siverio:2018}.

%
%
\section{Results} \label{sec:results}
For the purpose of showing the surges analyzed in this paper, left column of Figure~\ref{fig_inversions_map} contains radiance maps for the 
chromospheric \MgII\ k 2796.3~\AA{} line within the selected subFoV (solid box in Figure~\ref{fig_context}) during the four \IRIS\ 
raster scans. These maps are computed by averaging the intensity within $\pm 15 \,\mathrm{km\,s}^{-1}$ with respect to the core of the 
aforementioned line. In the panels, we have added contours to ease the location of the bulk of the surges, which can be recognized by their
dark and elongated structures composed by threads. The surges can be more extended spatially than the regions limited by these contours, since they refer only to the radiance maps in the center of \MgII\ k 2796.3~\AA{}. In the panels, it is also possible to distinguish the location of the UV bursts associated with the surges, 
specially in the third and fourth rasters, due to their enhanced brightness \citep[see][for details about these UV bursts]{Guglielmino:2019}. 
In the following, we focus on the observed surges by studying their chromospheric properties through the \MgII\ h\&k line (Section \ref{sec:mgii_analysis}), analyzing their transition region counterpart using the \OIV\ 1399.8~\AA{} and 1401.2~\AA{} lines (Section \ref{sec:oiv_analysis}), and comparing the observational results with the simulations
(Section \ref{sec:experiment_analysis}).

\begin{figure*}[!ht]
	\centering
	\includegraphics[width=1.00\textwidth]{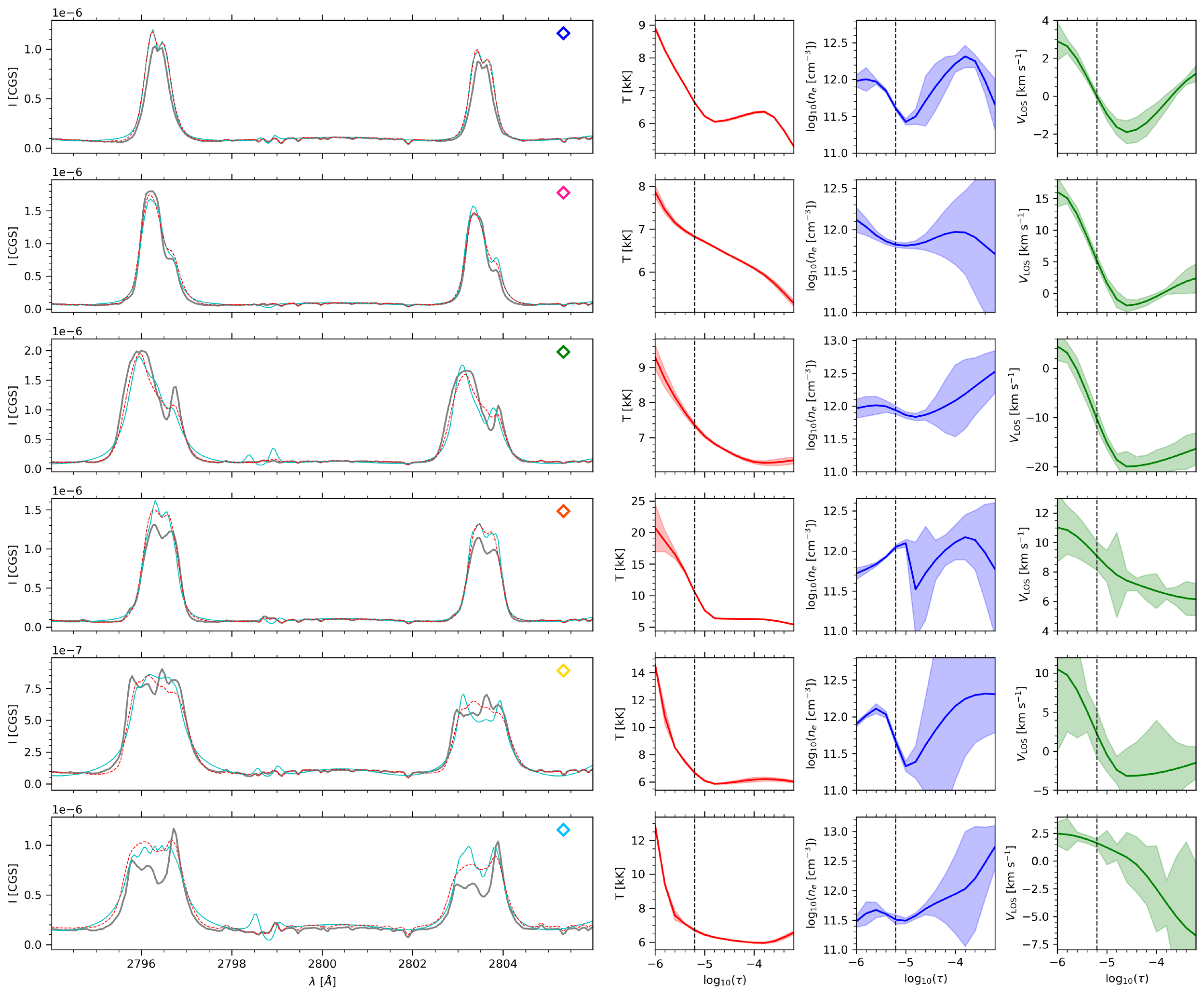}
	\caption{
	Spectral information and computed atmospheres for the six positions indicated with colored diamonds in Figure \ref{fig_inversions_map}. Left column: observed 
	\MgII\ h\&k profile (grey line), corresponding representative profile obtained with the \kmeans\ method (red dashed line), and the inverted profile using the STiC code (cyan line)
	in CGS units (erg cm$^{-2}$ s$^{-1}$ sr$^{-1}$ Hz$^{-1}$). Three rightmost columns: results from the inversions for $T$ (red), $n_e$ (blue), and \vlos\ (green) as functions of $\log_{10}(\tau)$ with their corresponding inversion uncertainties ($\pm \sigma$). The vertical dashed line in these plots at $\log_{10}(\tau)=-5.2$ indicates the optical depth shown in Figure \ref{fig_inversions_map}}.
	\label{fig_examples}
\end{figure*}

\subsection{Analysis of \texorpdfstring{\MgII}{Mg II} h\&k in the surges}\label{sec:mgii_analysis}

\subsubsection{\texorpdfstring{\MgII}{Mg II} h\&k clustering and profile examples}\label{sec:mgii_clustering}

The first step prior to analyze the \MgII\ h\&k spectra within surges is the clustering through the \kmeans\ method
(Section \ref{sec:kmeans}). In addition to identify representative profiles in the phenomenon of interest, this method allows us to noticeably 
reduce the number of \MgII\ h\&k profiles that would be in principle necessary to analyze. In our particular case, we have chosen $k=160$ 
for each raster, thus getting a total of 640 representative profiles that are later inverted. This implies a reduction of a factor 43.2 
with respect to the original total number of profiles (27648) within the \IRIS\ subFoV chosen for this paper. Second column of Figure 
\ref{fig_inversions_map} contains the label maps of the clusters for each raster. The labels are assigned to the clusters depending 
on the amount of profiles that compose each clusters: lower the label number, greater the number of profiles in the corresponding cluster. 
By ordering the labels, we can already discern that surges and their surroundings have clear different \MgII\ h\&k profiles 
from other regions.

In order to show the peculiarity of some of the surge profiles and to compare them with the representative ones obtained with the \kmeans, 
we have selected six different pixels in the surge of Raster 3 (colored diamonds in Figure~\ref{fig_inversions_map}), which are located in an 
elongated finger-like thread (blue); the left side of the bulk of the surge (pink); some particular regions at the boundaries of the ejection 
(orange, green and yellow); and the right side of the bulk of the surge (cyan). The later four locations are also used in the analysis of the 
\OIV\ lines (Section \ref{sec:oiv_analysis}). The \MgII\ h\&k spectral profiles for these six locations are plotted in the left column of Figure 
\ref{fig_examples} as grey lines. They show large diversity, with $k_{2V}$ and $h_{2V}$ peaks being greater than $k_{2R}$ and $h_{2R}$, respectively, 
excepting in the cyan location, suggesting different Doppler motions in the upper chromosphere \citep{Leenaarts:2013II}; $k_3$ and $h_3$ ranging 
from clearly visible valleys (as in the green location) to an almost flat behaviour (blue pixel); broad profiles with several peaks (yellow and cyan); 
and an interesting absence of emission in the \MgII\ triplet in all of them, which is contrarily observed in UV bursts \citep{Hong:2017}. 
In the panels, we also plot, as a red dashed line, the corresponding representative profiles obtained with the  \kmeans\ method. 
The representative profiles match quite well the observational data particularly for the blue, pink, orange and green locations,
which only show slightly differences in the intensity while the shape of the \MgII\ h\&k lines and the continuum are accurately represented. In the case of the
cyan and yellow locations, the continuum and the width of \MgII\ h\&k lines are properly captured with the representative profile; nonetheless, the shape of the peaks of the lines is less similar. We need to highlight that this discrepancy is a natural result expected from the \kmeans: there are
always some observed profiles that are assigned to the cluster with the most similar profiles but still they are not nicely represented. In our case, we have checked that these divergent
profiles represent a minority within the clusters (we have validated the good behaviour of the \kmeans\ algorithm by three different methods as mentioned in the Appendix \ref{sec:appendix}). Consequently, these small number of outliers have little impact on our results since we are interested in an overall view of the physical properties derived from inversions as discussed in the following section.

\subsubsection{Inversions of the representative \texorpdfstring{\MgII}{Mg II} h\&k profiles and statistics}\label{sec:mgii_inversion}
Once we have the representative profiles for each raster, we proceed with their corresponding inversions using the state-of-the-art 
STiC code with the scheme explained in Section \ref{sec:inversions}. Third, fourth and fifth columns of
Figure~\ref{fig_inversions_map} contain the results for the temperature, $T$; the electronic number density, $n_e$; and the LOS velocity, \vlos; respectively at $\log_{10}(\tau) = -5.2$. In the maps, it is possible to distinguish that the surges have some peculiarities in their physical properties such as they are mostly cooler and with smaller electron number density  than their surroundings. With respect to the velocity, the maps reveal that the bulk of the surge have values of a few km s$^{-1}$, while the largest values are located next to the edges of the contours of the finger-like threads (see Raster 3 and 6) reaching up to $\sim 11$ km s$^{-1}$. An animation of this figure is also available, showing the variation of the properties with the optical depth between $ -6.0 \le \log_{10}(\tau) \le -3.2$.

\begin{figure}[!ht]
	\centering
	\includegraphics[width=0.5\textwidth]{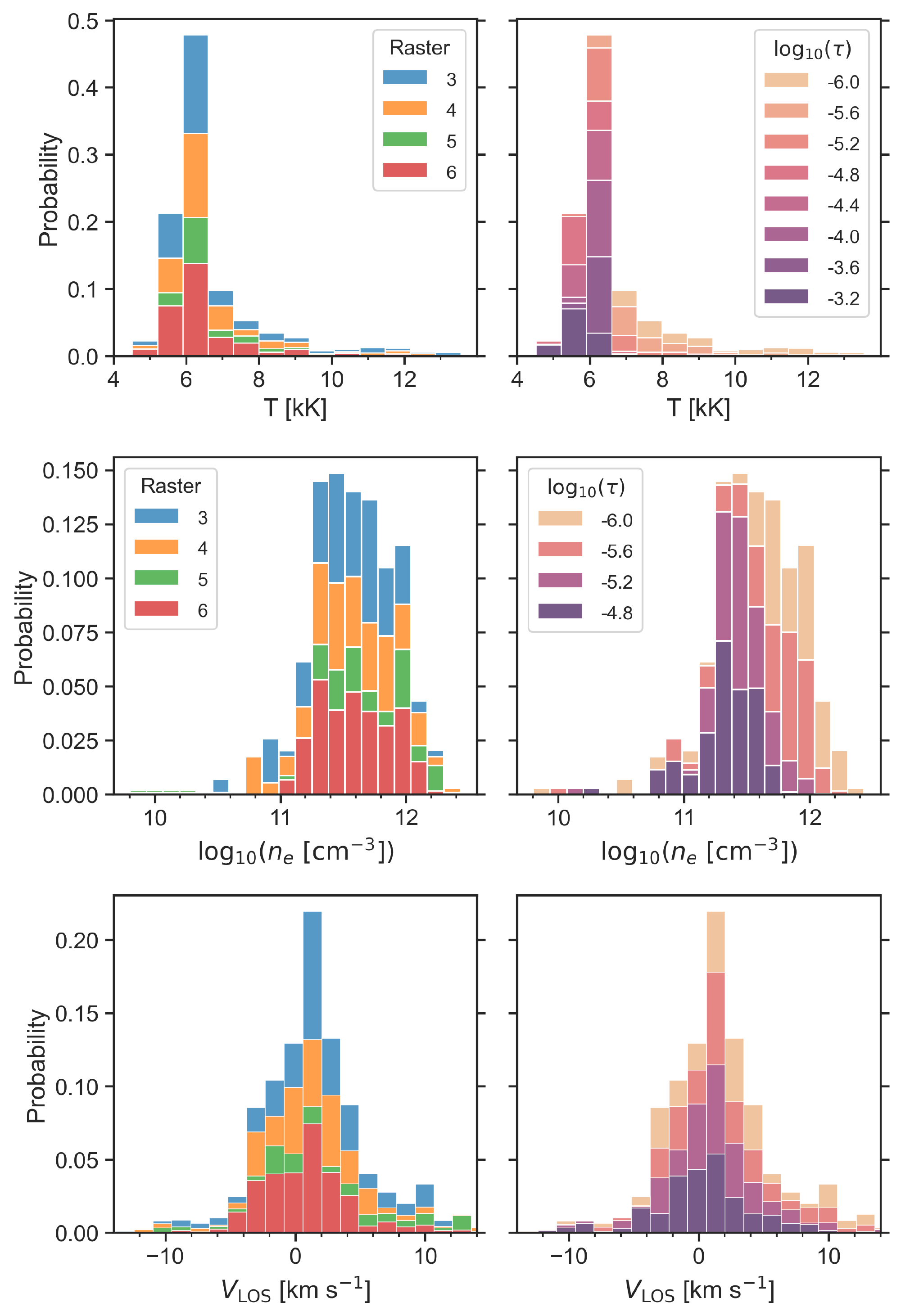}
	\caption{Statistical results for $T$ (top row), $n_e$ (middle row), and \vlos\ (bottom row) obtained within the contour that delimits the bulk of the different surges in Figure \ref{fig_inversions_map}. The statistics contain data from the optical depths where inversions are more reliable: from $\log_{10}(\tau) = -6.0$ to $-3.2$, for the temperature, and from $\log_{10}(\tau) = -6.0$ to $-4.8$ for the density and velocity. Left column: stacked histograms organized by raster showing that the different surges have similar properties. Right panel: stacked histograms organized by $\log_{10}(\tau)$ illustrating the variation of the physical parameters with the optical depth.}
	\label{fig_statistics}
\end{figure} 

In the panels of the left column of Figure \ref{fig_examples}, we have plotted the inverted profiles for the selected locations of Raster 3 as cyan lines to compare them with the representative profiles obtained with the \kmeans\ method. This way, we can get an idea about whether the inversion scheme used in this paper (Table \ref{table:inversions}) is adequate for such complex profiles. We have found that, in most cases, the inverted profiles mimic the representative profiles with high grade of detail and that the major discrepancies are in the complex profiles of the cyan and yellow locations. Right panel of Figure \ref{fig_examples} contains the corresponding representative model atmospheres extracted from the selected locations to illustrate the variation of $T$ (red), $n_e$ (blue), and \vlos (green) with $\log_{10}(\tau)$. For each physical quantity, we have also superimposed the inversion uncertainties ($\pm \sigma$ ). The uncertainties are small for the temperature in the whole range
of optical depths, while for the density, the smaller uncertainties are located upper in the atmosphere (more negative values of $\log_{10}(\tau)$). The latter is also true for the velocity in the blue, pink, green and cyan locations. We have inspected the behaviour of the uncertainties for all the results within the black contour that delimits the bulk of the different surges in Figure \ref{fig_inversions_map}, concluding that for the temperature, the accuracy of the inversions is excellent for all the optical depths within $\log_{10}(\tau) = -6.0$ and $-3.2$; for the density, the inversions results  can be rather acceptable between $\log_{10}(\tau) = -6.0$ and $-4.8$; and for the velocities, the range between $\log_{10}(\tau) = -6.0$ and $-4.8$ is also the best one generally.

To get a whole perspective of the general properties of the surges, we need to perform a statistical analysis. Figure \ref{fig_statistics} contains histograms for $T$ (top row), $n_e$ (middle row), and \vlos (bottom row), obtained within the black contour that delimits the bulk of the different surges in Figure \ref{fig_inversions_map}. To create these histograms, we consider the range of optical depths in which the uncertainties are smaller for each physical quantity. The histograms in the left column of the figure are stacked by rasters, that is, each bar in the chart represents a whole, and the length of the segments in the bar represents the contribution from each raster to the whole. Even though we have performed \kmeans\ individually for each raster with topologically different surges, the statistical distributions for temperature, density and velocity that we obtain after the inversions of the representative profiles are very similar: an important result that confers robustness to our method to characterize the properties of the surges. The characteristic properties of our surges are as follows: the most probable temperature is around $6$~kK, the electronic number density is mostly concentrated from $\sim 1.6 \times 10^{11}$ to $10^{12}$ cm$^{-3}$, and the LOS velocity is typically of a few km s$^{-1}$. The right column of Figure \ref{fig_statistics} shows the same statistics but stacked by optical depth, highlighting the variation of the parameters with the optical depth. In this case, cooler plasma with smaller electron number density in the surges is located at less negative values of the optical depth, that is, deeper in the atmosphere.

\subsection{Analysis of \texorpdfstring{\OIV}{O IV} in the surges}\label{sec:oiv_analysis}

\begin{figure*}[!ht]
	\centering
	\includegraphics[width=1.0\textwidth]{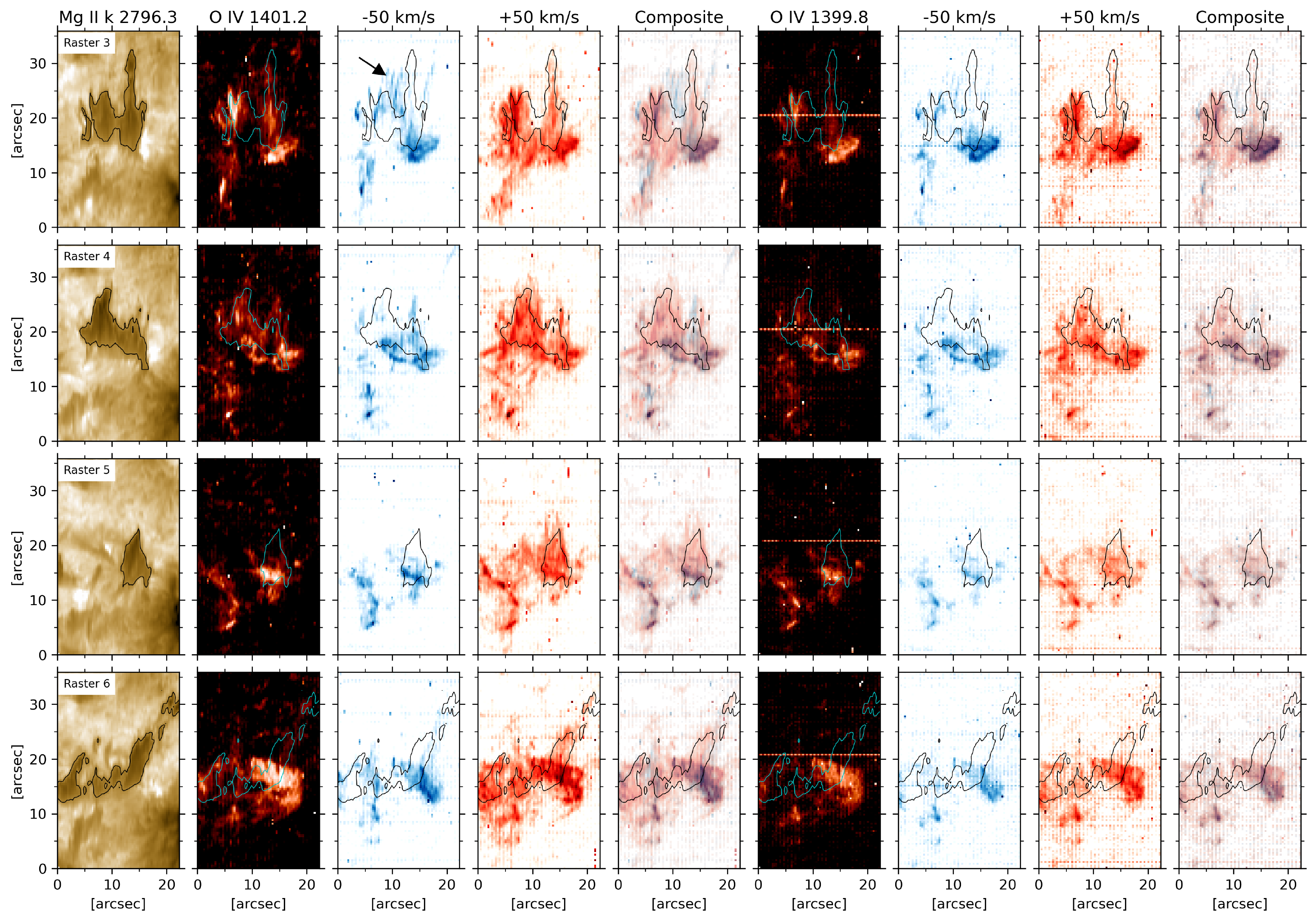}
	\caption{Radiance maps showing the \OIV\ emissivity of the surges.
	From left to right: radiance maps in the core of the \MgII\ k 2796.3~\AA{} line (first column); 
	in the core of the \OIV\ 1401.2~\AA{} line (second column); 
	in the blue and red wings at $-50$ and $50 \,\mathrm{km\,s}^{-1}$ of \OIV\ 1401.2~\AA{} (third and fourth columns, respectively); 
	and composite image of the blue and red wing radiance maps for \OIV\ 1401.2~\AA{} (fifth column). 
	Equivalent maps are plotted for the \OIV\ 1399.8~\AA{} line (sixth-ninth columns).
	The contours in all the panels delimit the bulk of the surges visible in the \MgII\ k 2796.3~\AA{} radiance maps. The black arrow
	in the third panel of the first row indicates the location of enhanced \OIV\ emission associated with finger-like threads of a surge.
	The \OIV\ maps show a horizontal strip of hot pixels that does not affect our findings.} 
	\label{fig_oivmaps}
\end{figure*}

\subsubsection{Enhanced \texorpdfstring{\OIV}{O IV} emissivity within the surge}\label{sec:oiv_emissivity}
One of the most striking results is the finding, for the first time, of emission in the \OIV\ 1401.2~\AA{} and \OIV\ 1399.8~\AA{} lines 
in surges. To illustrate this fact, Figure~\ref{fig_oivmaps} contains the \MgII\ k 2796.3~\AA{} line radiance maps for the four 
rasters (first column), to show the context, and the corresponding radiance maps for \OIV\ 1401.2~\AA{} (second column) and for \OIV\ 1399.8~\AA{} (sixth column). 
These radiance maps are computed by averaging the intensity within $\pm 15 \,\mathrm{km\,s}^{-1}$ 
with respect to the core of each line. In addition, for both \OIV\ lines, we compute maps of integrated emission in the blue wing at 
$- 50 \,\mathrm{km\,s}^{-1}$ (third/seventh column of the figure) and in the red wing at $+ 50 \,\mathrm{km\,s}^{-1}$ (fourth/eighth column). 
Differences in the location of the emitting regions between the wings are determined from the composite maps (fifth/ninth column of the
image). To ease the identification, all the panels show contours that delimit the bulk of the surges visible in the \MgII\ k 2796.3~\AA{} radiance maps.
The maps, specially the radiance ones of \OIV\ 1399.8~\AA{}, contain a horizontal strip of hot pixels; however, this does not affect our results.

Inspecting the image, we can see that both \OIV\ 1401.2~\AA{} and \OIV\ 1399.8~\AA{}  radiance maps at the center of the line (second and sixth columns) 
look quite similar, although the surges appear to have a larger extension in \OIV\ 1401.2~\AA{} than in \OIV\ 1399.8~\AA{}, likely due to the fact that 
the former is a stronger line. The location of the brightest \OIV\ regions can be found within the bulk of the surges and/or in their boundaries following the threads of the surges.
This is particularly evident in the third and fourth rasters. The corresponding blue/red wing and composite maps also provide evidences about the transition region counterpart 
of the surges in these weak  \OIV\ lines. For instance, in the third panel of the first row of the image, a black arrow indicates the location of enhanced \OIV\ emission associated 
with finger-like threads of the surge. In addition, the maps also reveal a peculiar behavior in these threads: while several threads are observed co-located both in the blue and red wings of the \OIV\ lines, some others show a significant shift in the position where they are detected in the blue and red wings of those lines. This asymmetry is similar to the noticed for \ion{Mg}{ii}~k and \SiIV\ 1402.8~\AA{} lines by \citet{Guglielmino:2019} and it could be an evidence of helical/rotational motions typically associated with surges.

\subsubsection{Spectral analysis and density diagnostics}\label{sec:oiv_diagnostics}
\begin{figure*}[!h]
	\centering
	\includegraphics[width=1.0\textwidth]{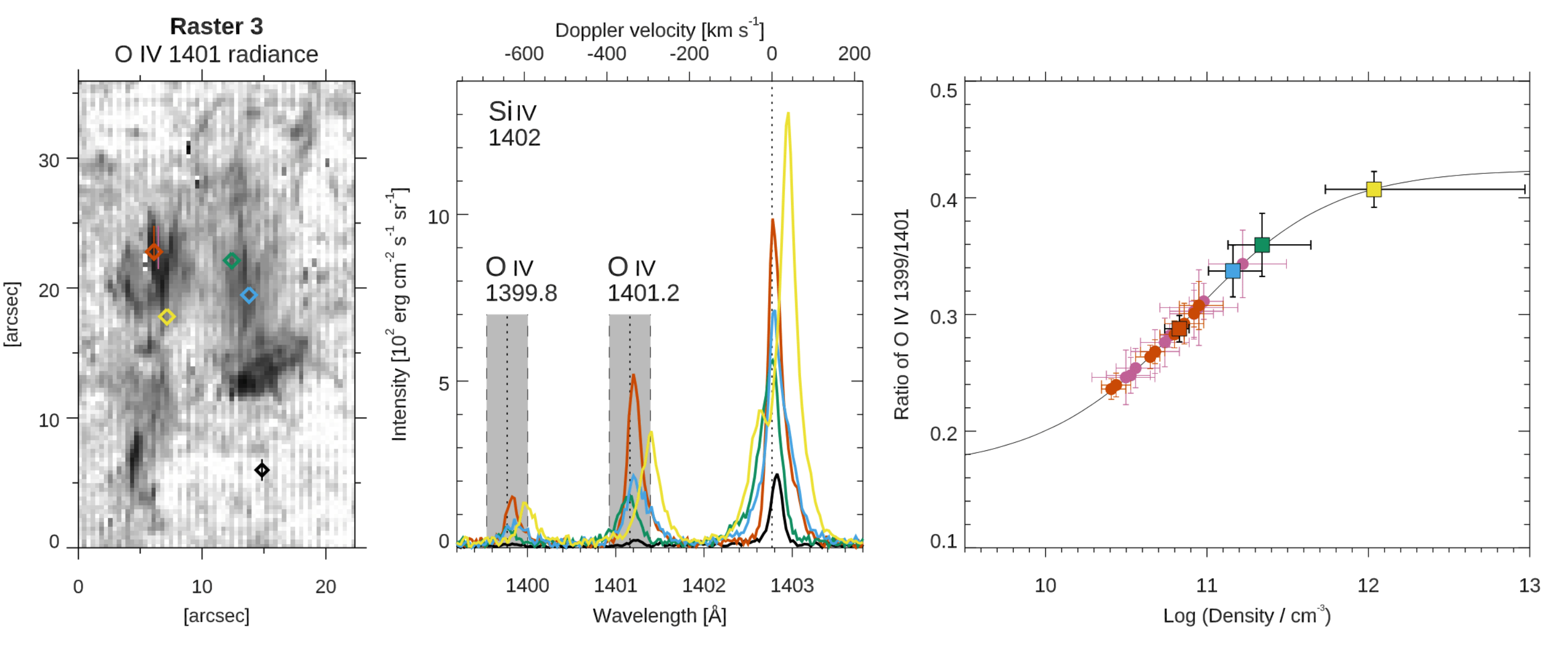}
	\caption{\OIV\ spectral analysis and density diagnostics. Left: \OIV\ 1401.2~\AA{} radiance map (reversed color scale) for the third raster with superimposed colored diamonds (orange, green, blue, and yellow) and colored segments (orange and magenta) where the spectra are analyzed. For reference, the black diamond illustrates a quiet-Sun area with a vertical segment of six pixels in which the quiet-Sun spectrum has been averaged. Middle: FUV spectra in the \SiIV\ 1402 \AA{} passband for the positions indicated in the left map with diamonds.
	Grey shaded areas delimit spectral regions at $\pm50 \,\mathrm{km\,s}^{-1}$ encompassing \OIV\ 1399.8~\AA{} and \OIV\ 1401.2~\AA{} lines. 
	Right panel: Results from the density diagnostics using the \OIV\ 1399.8/1401.2 ratio for the surge pixels selected in the left panel.
	} \label{fig_density_diagnostics}
\end{figure*}

To know how the \OIV\ spectra related to the surges look like, left panel of Figure~\ref{fig_density_diagnostics} contains the \OIV\ 1401.2~\AA{}
radiance map with reversed color scale for the third raster. From our previous choice of locations in Figure \ref{fig_inversions_map},
we have selected the pixels with strong intensity in both \OIV\ lines (orange, yellow, green, and cyan diamonds). In addition, we also 
consider some other pixels along a thread of the surge, approximately located at $x=6 \arcsec$ and $y\in [20\arcsec,25\arcsec] $, 
that have the greatest \OIV\ intensity (orange and magenta segments). The black diamond in the map indicates a quiet-Sun area with a 
vertical segment of six pixels in which the spectrum has been averaged for comparison purposes. The corresponding surge and quiet-Sun 
spectra are shown in the middle panel of Figure~\ref{fig_density_diagnostics} using the same colors as for the diamonds. In this panel, 
we can see that the \OIV\ intensity associated with the surge is a factor between 7 and 15, for the \OIV\ 1399.8~\AA{} line, and between 
12 and 35, for the \OIV\ 1401.2~\AA{} line, brighter than the quiet-Sun average profile: surges have a transition region counterpart even 
in the weakest FUV lines. This is moreover very different from UV bursts, where the \OIV\ 1401.2~\AA{} line is very weak in comparison to 
\SiIV\ 1402.8, and the \OIV\ 1399.8~\AA{} line is absent \citep{Peter:2014h,Young:2018}. With respect to the \SiIV\ 1402.8 intensity, the 
ratio between the surge and the quiet-Sun region intensities ranges from ~2.5 to 6.5, which agrees with the values previously found in 
\IRIS\ observations (from 2 to 5) by \citealp{Nobrega-Siverio:2017}. 

The simultaneous finding of \OIV\ 1399.8~\AA{} and 1401.2~\AA{} allows us to estimate the electron density in the 
transition region of the surges. Applying the density diagnostics method described in Section \ref{sec:density_diagnostics} for the selected 
pixels (diamonds and vertical segments in the left panel of Figure~\ref{fig_density_diagnostics}), we obtain a density in the range of 
$\approx 2.5 \times 10^{10} - 10^{12} \,\mathrm{cm}^{-3}$ as shown in the right panel of Figure~\ref{fig_oivmaps}, where the yellow pixel 
lies at the high-density limit of the diagnostic method. 

The next question to address is whether it is possible to establish a connection between the physical quantities that are derived for the transition region
from the \OIV\ analysis and the ones obtained for the mid- and low-chromosphere in Section \ref{sec:mgii_inversion} from the \MgII\ h\&k lines. The green location, for instance,
shows an \OIV\ spectrum (middle panel of Figure~\ref{fig_density_diagnostics}) that appears to be taken in a pixel almost at rest, while the 
electron density (right panel of Figure~\ref{fig_density_diagnostics}) is around $2\times 10^{11}\,\mathrm{cm}^{-3}$. For this location, the \MgII\ h\&k inversions 
(Figure~\ref{fig_examples}) show that, at $\log_{10}(\tau)=-6$, the velocity is small ($\sim 1\pm 1$ km s$^{-1}$),  which is consistent 
with the Doppler shift found in \OIV, and the density is close to $10^{12} \,\mathrm{cm}^{-3}$ and decreases towards higher layers in the atmosphere, which also 
seems to be consistent if we assume that the \OIV\ information comes from an upper and more rarefied layer. The orange  
pixel has an \OIV\ spectrum exhibiting a slight redshift of about $10 \,\mathrm{km\,s}^{-1}$ and a density of $\approx 8 \times 10^{10} \mathrm{cm}^{-3}$.
This also appears to be in agreement with the results from the inversions, which show redshift velocities of $\sim 11\pm 3 \,\mathrm{km\,s}^{-1}$ and
a density of $5 \times 10^{11} \,\mathrm{cm}^{-3}$ rapidly decreasing upwards. The yellow and cyan locations show both redshifts in \OIV, being
the yellow location the one with larger velocities and more electron density. This general behavior is also found in the inversions, although the comparison
is more complicated owing to the uncertainties involved in these two locations. In the following section, we compare these results with those
from numerical simulations.

\subsection{Comparison between observations and numerical experiments}\label{sec:experiment_analysis}

\begin{figure*}[!h]
	\centering
	\includegraphics[width=1\textwidth]{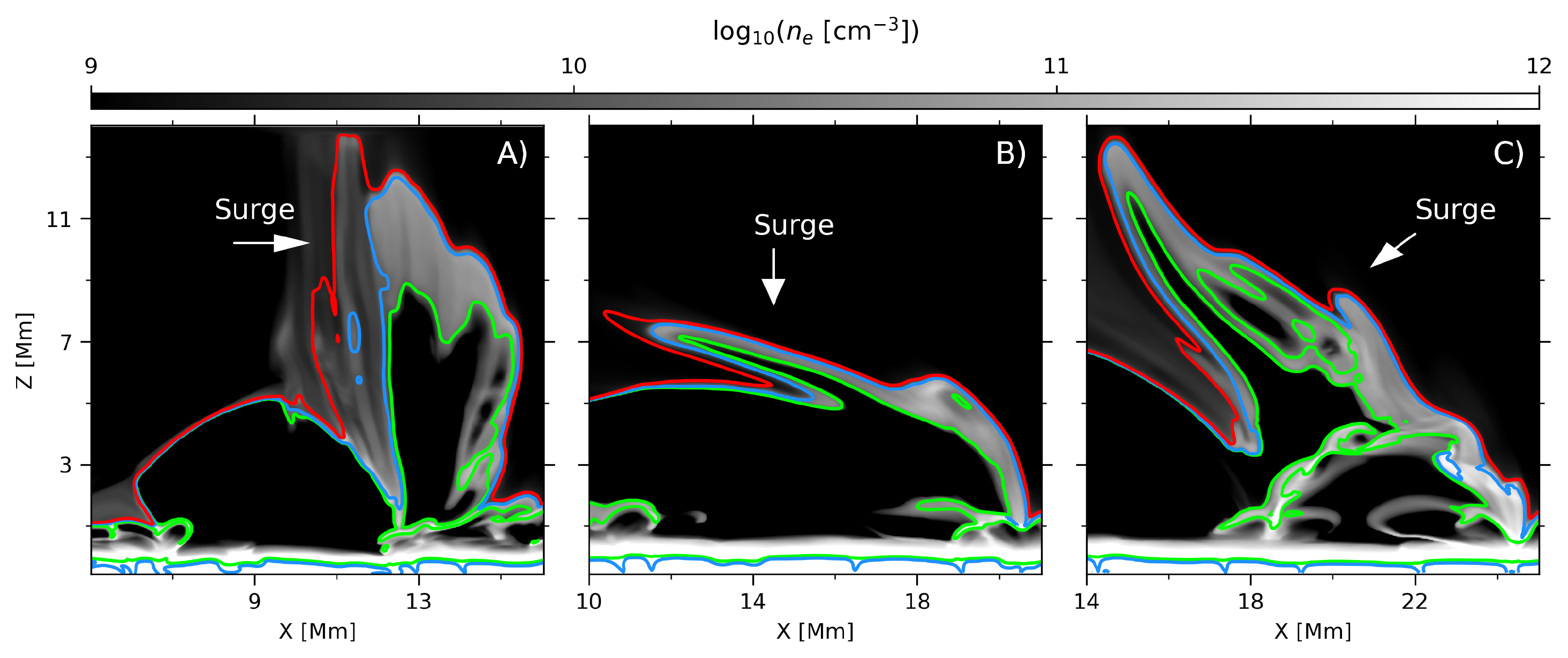}
	\caption{Electron number density for three different simulated surges. Contours of temperature are superimposed for $T=6$~kK (green), $T=10$~kK (blue) and
	$T=200$~kK (red). Panel A: surge from the \cite{Nobrega-Siverio:2016} numerical experiment. Panels B and C: surges from the simulations by \cite{Nobrega-Siverio:2017,Nobrega-Siverio:2018}.} \label{fig_simulations}
\end{figure*}

To get a joint perspective between observations and theory, Figure \ref{fig_simulations} contains electron number density maps 
for three different 2.5D numerical experiments at the time when surges are at their apex. The ejections are discernible as elongated 
structures and their locations are indicated with arrows. The three surges show finger like structures similar to the observed ones.
It is also possible to distinguish, specially in the left panel, that $n_e$ shows variations within the surge that look like threads. 
Temperature contours are superimposed for $T=6$~kK (green), $T=10$~kK (blue) and $T=200$~kK (red), highlighting the multi-thermal
structure of these phenomena. Focusing on the location of these contours, we can see that in some regions of the surge there are large 
temperature gradients with cool chromospheric plasma being close to the transition region plasma (where \OIV\ emission is originated), while 
in other regions, the transition region of the surge is far away (from hundred km up to a few Mm) from the cool core. 
This may explain why the location of the observed brightest \OIV\ regions in Figure \ref{fig_oivmaps} can be found within the bulk of the 
surges and/or in their boundaries following the threads of the surges. Around the $T=6$~kK contour, the surges have a clear core with 
small electron number density. The number density increases from the core outwards until reaching 
transition region, where it decreases again. Table \ref{table:experiments} contains the averaged $n_e$ in the temperature contours for the 
three experiments as a zero order approximation to compare with the observations. By doing this, we obtain number densities that are 
lower than those derived from observations. This means that, although qualitatively the experiments help us to understand
the observed features, for a more accurate comparison we would need to include nonequilibrium effects to properly model the chromosphere 
and synthesize \MgII\ h\&k lines in the experiments (see Discussion).

\begin{table}
\centering
\begin{tabular}{ c c c c}
 \hline
  \hline
 \rule{0pt}{2.5ex}
$\log_{10}(n_e$  [cm$^{-3}$]) & Exp. A & Exp. B & Exp. C \\
 \hline
\rule{0pt}{2.5ex} 
 at $T=6 \mathrm{kK}$  & $10.6$ &   $10.8$ &   $10.5$ \\
 \rule{0pt}{2.5ex}
 at $T=10 \mathrm{kK}$   & $11.4$ &   $12.7$ &   $12.5$ \\
  \rule{0pt}{2.5ex}
 at $T=200 \mathrm{kK}$  & $9.6$    &   $9.3$  &   $9.5$ \\ 
\hline 
\end{tabular}
\caption{Averaged electron number density in the temperature contours of the numerical experiments shown in Figure \ref{fig_simulations}.\label{table:experiments}}
\centering
\end{table}

%
%
\section{Discussion and conclusions} \label{sec:discussion}
In this paper, we have used observations from the \IRIS\ satellite to study an episode of recurrent surges that appear associated with 
UV bursts in Active Region (AR) NOAA 12529 on 2016 April. This is the first study of surges that characterizes their fundamental
properties using the \MgII\ h\&k as well as the \OIV\ 1399.8 and 1401.1~\AA{} spectral lines, combining to that end different methods 
such as the \kmeans\ algorithm, inversions with the STiC code, and density diagnostics. In addition, we have employed numerical
experiments using the Bifrost code to get a qualitative comparison between observations and theory. In the following, we summarize 
and discuss the main findings of this paper.

In Section \ref{sec:mgii_analysis}, we have analyzed the near-UV spectra
of surges through the \MgII\ h\&k line, focusing on the properties of the surges as they have been mainly unexplored.
    In Section \ref{sec:mgii_clustering}, we have used the \kmeans\ method in each of the four selected rasters to find \MgII\ h\&k representative profiles in our observation and invert them. Thanks to this method, we reduce the number of profiles to invert by a factor 43.2, thus alleviating the computational and analysis effort, as well as being able to discern regions with peculiar profiles, like in the case of surges and UV bursts. The \kmeans\ algorithm is not the only approach for this kind of studies. For instance, \cite{Verma:2021} employ $t$-distributed stochastic neighbor embedding (t-SNE), another machine learning algorithm, to classify the H$\alpha$ spectra with the aim of separating different regions to perform cloud model inversions. 
    In Section \ref{sec:mgii_inversion}, we have inverted our representative profiles from the \kmeans\ finding that the surges have
    their most probable temperature around $6$~kK, electronic number densities mostly concentrated from $\sim 1.6 \times 10^{11}$ to $10^{12}$ cm$^{-3}$, and LOS velocities of a few km s$^{-1}$. In addition, we have also showed that the statistical distributions of these properties are very similar for the different surges, meaning that these ejections can be well constrained in terms of their physical quantities. The uncertainties analysis show that the values of temperature for the surges are very reliable for optical depths between $\log_{10}(\tau)=-6.0$ and $-3.2$, the density values can only be considered between $\log_{10}(\tau)=-6.0$ to $-4.8$, while the velocities are generally acceptable also between $\log_{10}(\tau)=-6.0$ to $-4.8$, although it is preferred to examine them individually pixel by pixel.
    These results have allowed us not only to gain new insights of fundamental properties of the surges, but also to constitute the first steps to characterize the surge spectral profiles in the main chromospheric lines in which they are observed. To the best of our knowledge, there is only one paper \citep{Kejun:1996} in which physical parameters of surges are computed through a two-component cloud model using \Halpha, obtaining $T=8.5$~kK and $n_e=1.8\times10^{10}$ cm$^{-3}$. More recently, inverting the \ion{Ca}{II} 8542~\AA\ line using the NLTE code NICOLE, \cite{Kuridze:2021} obtain the electron density values of a large \textit{spicule} off the limb. Their results show that $n_e$ ranges from $2.5\times10^{13}$ cm$^{-3}$, at the \textit{spicule} bottom, to $\sim 2\times10^{10}$ cm$^{-3}$, at its top. They argue that the density obtained is more akin to surges rather than typical spicules. In our case, the results from the inversion reveal a more constrained range for the electronic number density.

In Section \ref{sec:oiv_analysis}, we have focused on the transition region counterpart of surges through the FUV spectra. 
Thanks to the relatively long exposure time used in the \IRIS\ sequence (18 s for FUV, thus ensuring good signal-to-noise 
ratio for faint lines in the transition region), in Section \ref{sec:oiv_emissivity} we have been able to find emission in the 
\OIV\ 1399.8~\AA{} and 1401.1~\AA{} lines related to the surges. This finding is relevant due to the following reasons:
    i) this is the first observational report of \OIV\ 1399.8~\AA{} and 1401.1~\AA{} related to surges, in fact, previous FUV 
    analysis could not detect any \OIV\ emission because the exposure time was too short \citep{Nobrega-Siverio:2017}, 
    or because the focus was on other lines \citep{Guglielmino:2019};
    ii) it clearly demonstrates that surges have a transition region counterpart even in the weakest FUV lines (the oscillator 
    strength, which is proportional to the line strength, of 1401.1~\AA{} is $5.1\times 10^{-7}$ and the one for \OIV\ 1399.8~\AA{} is 
    $4.3\times10^{-7}$, which is a 16\% smaller);
    iii) it gives observational support to the theoretical findings by \cite{Nobrega-Siverio:2018}, which showed that the surges can 
    have enhanced emissivity in lines like \OIV\ 1401.1~\AA{}. 
%
%
We have also found that surges have threads with important blue- and redshifts that can appear spatially displaced (see composite maps 
of Figure \ref{fig_oivmaps}). This could be an indication of helical/rotational motions in the transition region counterpart of the surges 
during their untwisting and ejection. Similar asymmetry in the locations of the blue- and redshifts within surges is also found in the 
\ion{Mg}{ii}~k and \SiIV\ 1402~\AA{} lines by \citet{Guglielmino:2019}, and they join to the further indications of rotational motions 
associated with surges described previously in the literature \citep[see, e.g.,][]{canfield1996,Jibben2004,Jiang2007,Bong:2014}. 
In Section \ref{sec:oiv_diagnostics}, we have selected the brightest \OIV\ pixels in the surge to perform density diagnostics, 
finding that the electron number density ranges from, 
approximately, $2.5 \times 10^{10}$ to 10$^{12}$ cm$^{-3}$. The physical quantities 
obtained from \OIV\ seem to be consistent with those derived from \MgII\ h\&k inversions for these locations. However, one has
to be cautious: the \OIV\ intensity we observe can come from the integration along the LOS of different regions of 
the surge, as shown in numerical experiments by \cite{Nobrega-Siverio:2018}, so the inferred electron density could be just a weighted 
mean average electron density along the line of sight. In addition, nonequilibrium ionization can be relevant for these emitting layers, 
and density diagnostics may be affected by an order of magnitude \citep{Olluri:2013fu}. A possible way of better constrain the
physical properties of surges in the future is using inversions combining the information from the chromosphere and transition region 
in a similar fashion as \cite{Vissers:2019} carried out for UV bursts and Ellermam bombs.

In Section \ref{sec:experiment_analysis}, we qualitatively compare three simulations with the observations finding 
similarities in terms of the topology that may also explain the location of the observed brightest \OIV\ regions with respect
to the bulk/core of the surges. \cite{Nobrega-Siverio:2016} show that the bulk of their simulated surge (population A in that paper) 
is quite concentrated around $6$~kK, which agrees with the values from the inversions we find here. The electron
number density, instead, seems to be lower in the simulations. This could be related to the lack of nonequilibrium ionization in 
these experiments. \cite{Nobrega-Siverio:2020a} demonstrate that the LTE assumption 
substantially underestimates the ionization fraction in most of the emerged region, which could lead to lower electron number densities
in the subsequent surge. Therefore, numerical simulations of surges including the ionization out-of-equilibrium of hydrogen seems mandatory
to match the physical properties derived from observations.

The combination of methods and results obtained in this paper open new possibilities for the analysis and diagnostics of surges. 
For instance, our inverted representative profiles and their corresponding representative model atmospheres will be included in the 
\IRIS$^2$ \citep{Sainz-Dalda:2019} database, improving its current capabilities to invert eruptive phenomena. These results may 
facilitate building automatic detection algorithms for these key ejections in observations. The simultaneous inversions of different 
spectral lines in which surges are observed will also allow us to compare and better constrain numerical experiments. In this vein, in 
the near future we expect to exploit the existing high-quality data by the Swedish 1-m Solar Telescope (SST) to perform detailed 
characterization of surges in other lines such as  \CaII\ H\&K and \CaII\ 8542 $\angstrom$ \citep[see, e.g., the publicly available 
co-aligned IRIS and SST datasets by][]{Rouppe:2020} . In addition, future observations from, for instance, the SPICE instrument on 
board the Solar Orbiter mission \citep{Spice:2020} could help us to continue exploring the upper chromosphere and transition region 
of these important phenomena through lines such as \ion{H}{I} 1025.7~\AA{}, $\log_{10}(T)=4$~K; \ion{C}{III} 977.0~\AA{}, $\log_{10}(T)=4.5$~K; and 
\ion{O}{VI} 1031.9~\AA{}, $\log_{10}(T)=5.5$~K. In the same way, the 4-m diameter Daniel K. Inouye Solar Telescope \citep[DKIST,][]{Rimmele:2020} 
could offer potential diagnostics at upper chromosphere and transition region temperatures of surges, for example, in the \ion{He}{I} 
D3 and 10830~\AA{} lines.

%
%
\appendix
\section{Processing for the \texorpdfstring{\kmeans}{k-means} algorithm}\label{sec:appendix}
To obtain meaningful results out of the \kmeans\ clustering, 
it is necessary to process the observational data properly. To that end, 
we have carried out the following steps.

\begin{figure}[!ht]
\centering
\includegraphics[width=0.5\textwidth]{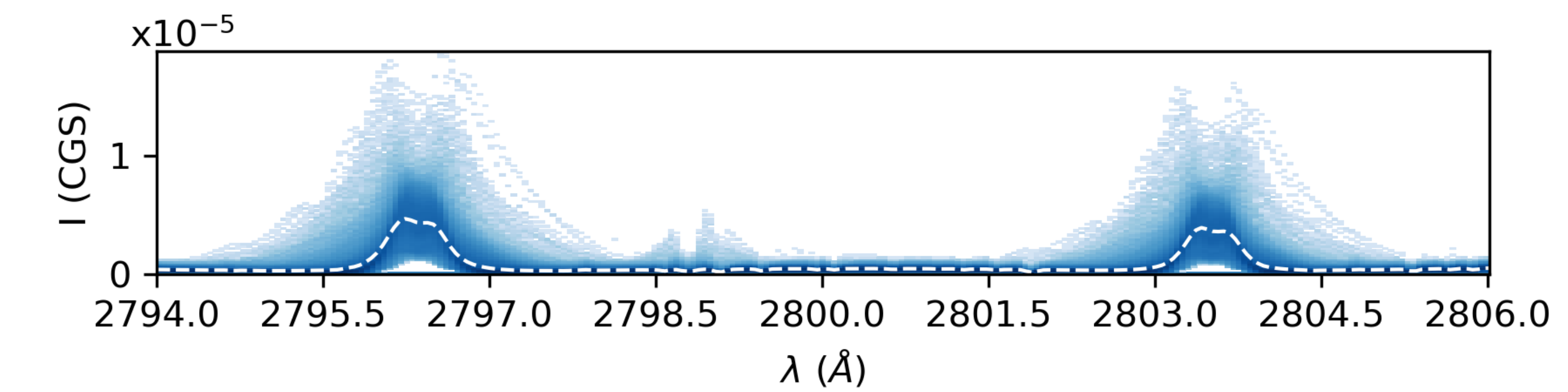}
\includegraphics[width=0.5\textwidth]{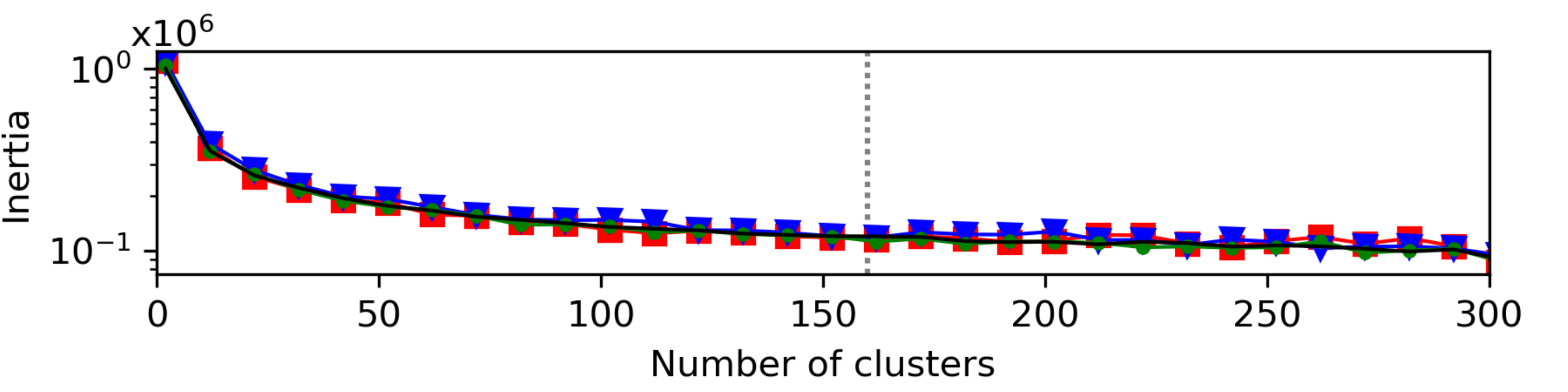}
\caption{Data information for the \kmeans\ clustering. Top: Joint-PDF of all the
calibrated \MgII\ h\&k line profiles in CGS units (erg cm$^{-2}$ s$^{-1}$ sr$^{-1}$ Hz$^{-1}$) within the subFoV of Figure \ref{fig_context} for the four rasters. 
The corresponding average profile is shown as a dashed line. 
Bottom: Behavior of the inertia
 of the \kmeans\ method with increasing clusters to choose a suitable number through the \textit{elbow technique} for raster 1 (red square), raster 2 (blue triangle), raster 3 (green circle), raster 4 (black line). Our cluster choice for all the rasters, $k=160$, is indicated by the vertical line.}
\label{fig:appendix}
\end{figure}

\begin{enumerate}

\item Selection of the domain. 
We have chosen the subFoV of Figure \ref{fig_context} 
for our \kmeans\ clustering, thus reducing the number of profiles to 
be grouped that are not the aim of this paper. 

\item Removing bad data. 
Any profile affected by data loss or cosmic rays 
is replaced by zero values.
This is helpful as a benchmark: our data has 327 profiles 
affected and the \kmeans\ group them together in one single cluster,
thus being able to easily mask them in the figures of the paper.

\item Radiometric calibration.
We have performed radiometric calibration\footnote{See the documentation
about \IRIS\ Level 2 data calibration in \url{https://iris.lmsal.com/iris2/iris2_chapter04_01.html\#calibration-of-iris-level-2-data}} of the observational data to convert them into physical units. 
This step is essential to use the results obtained by the \kmeans\ method as  
input for the inversions. 

\item Cropping the spectra. 
We have cropped the spectra to the range between
2794 to 2806~\AA{}, which is sampled with 237 points. 
This is enough to contain the whole \MgII\ h\&k line.
Top panel of Figure \ref{fig:appendix} shows the distribution
of all the calibrated profiles through a joint-PDF plot as well as the average profile (dashed line).

\item Standardization of individual features. This step is a common requirement for machine learning 
estimators to avoid bad behavior if the features are not more or less normally distributed.
Our features are the wavelength positions, so for all the 237 positions
between 2794 to 2806~\AA{}, we have carried out a standardization of the 
intensity.
\end{enumerate}

After processing all the data, we have performed the \kmeans\ method using the tools available on Scikit-learn 
\citep{Pedregosa:2011} and selecting 160 clusters for each of the four rasters. 
The initialization of clusters is repeated 100 times, keeping the best output of the consecutive runs in terms of inertia. We have concluded that $k=160$ is an appropriate number of clusters through three different methods: visual comparison of the representative profiles
with the observed profiles belonging to a given cluster through joint-PDFs; using the benchmark of the bad pixels mentioned before; and 
checking the behavior of the inertia with the number of clusters (the \textit{elbow technique} shown in bottom panel of Figure \ref{fig:appendix}). In the figure, the regime $k\gtrsim 80$ shows little variation of the inertia for the four rasters. Since our
purpose are inversions and not finding the minimum number of clusters in which
the data can be grouped, we can take any $k$ in this regime because
the inversions will give very similar results.

%
%
\begin{acknowledgements}
This research is supported by the Research Council of Norway through its Centres of Excellence scheme, project number 262622, as well as through the
Synergy Grant number 810218 (ERC-2018-SyG) of the European Research Council, in addition to the project PGC2018-095832-B-I00 of the the Spanish Ministry of Science, Innovation and Universities, and
the ISSI Bern support for the team \textit{Unraveling surges: a joint perspective from numerical
models, observations, and machine learning}. This research has received funding from the European Commission's Seventh 
Framework Programme under the grant agreement no.~312495 (SOLARNET project) 
and from the European Union's Horizon 2020 research and innovation programme 
under the grant agreements no.~739500 (PRE-EST project) and no.~824135 
(SOLARNET project). S.L.G. acknowledges support by the Italian MIUR-PRIN grant 
2017APKP7T on \textit{Circumterrestrial Environment: Impact of Sun-Earth 
Interaction} and support by the Italian Space Agency (ASI) under contract 
2021-12-HH.0 to the co-financing INAF for the Italian contribution to the 
Solar-C EUVST preparatory science programme. A.S.D. contribution to this 
research is supported by NASA under contract NNG09FA40C (IRIS).
\end{acknowledgements}

%
%
\bibliographystyle{aa}
\bibliography{collectionbib}

\end{document}